\documentclass[aps,prl,showpacs,twocolumn,superscriptaddress,groupedaddress]{revtex4}  
\usepackage{graphicx,dblfloatfix}  
\usepackage{dcolumn}   
\usepackage{bm}        
\usepackage{amssymb}   
\usepackage{amsmath}
\usepackage[english]{babel}

\begin{document}

\title{Chiral molecules on curved colloidal membranes.}
\author{Sk Ashif Akram$^1$, Arabinda Behera$^1$,  Prerna Sharma$^2$ and Anirban Sain$^{1,*}$\\
\textit{$^1$ Department of Physics, Indian Institute of Technology-Bombay, Powai, Mumbai 400076, India, $^2$ Indian Institute of Science, Bangalore, 560012, India. }}
\date{\today}

\begin{abstract}
Colloidal membranes, self assembled monolayers of aligned rod like molecules, offer a template for designing membranes with definite shapes 
and curvature, and possibly new functionalities in the future. 
Often the constituent rods, due to their molecular chirality, are tilted with 
respect to the membrane normal. Spatial patterns of this tilt on 
curved membranes result from a competition among depletion forces,
nematic interaction, molecular chirality and boundary effects. 
We present a covariant theory for the tilt pattern on minimal surfaces, 
like helicoids and catenoids, which have been generated in the laboratory 
only recently. 
We predict several non-uniform tilt patterns, some of which are
consistent with experimental observations and some, which are yet to be 
discovered. 
\end{abstract}

\maketitle

\section{Introduction}
Rod like chiral molecules are used as a model system for studying how molecular chirality at microscopic scale generates  self assembled structures at mesoscopic scales, often with multiple polymorphic forms 
\cite{chiralstrucPNAS2001}. Ability to switch between these forms  
require understanding of how these structures respond to physical 
factors like temperature, chemical environment, composition of the constitutent rods and degree of their molecular chirality.
In the presence of depletion agents, like nonadsorbing polymers,
chiral or achiral rod like molecules (e.g., {\it fd} virus particles) 
can align and self assemble 
into monolayers that are one rod-length-thick and are often called 
colloidal membranes \cite{memdogic2009, memdogic2010,prerna}.   Unlike  
bilayer phopholipid membranes which typically form closed vesicles, colloidal membranes exhibit various open structures 
like flat circular discs, twisted ribbons \cite{Gibaud}, saddles and catenoids \cite{apsabstract}, in response to change of temperature, 
depletant concentration, or by mixing rods of different chirality, 
or rods of different lengths but same chirality. 

Here we focus on colloidal membranes with curved surfaces. 
While flat circular discs are the most stable structures, a disc can be 
transformed into a twisted ribbon (helicoid) by lowering temperature 
\cite{Gibaud} or by applying stretching force \cite{powers1}. Lowering 
temperature is known to increase molecular chirality which in turn
lowers the effective line tension at the membrane edges. This leads 
to proliferation of boundaries \cite{Gibaud}, and thus a switch from 
disc to a twisted ribbon. On the other hand 
mixing rods of two different lengths (but same chirality) generates 
membrane with negative gaussian curvatures as in saddles and 
catenoids \cite{apsabstract}. This has been interpreted as the result of a 
positive gaussian curvature modulus ($\kappa_G$) \cite{gibaud2017, apsabstract}. Interestingly, all these structures, namely, helicoid, catenoid are 
examples of minimal surfaces with zero mean curvature. That implies 
a large  bending modulus ($\kappa_B$) for these membranes. 
Large bending energy cost not only favours generation of minimal 
surfaces, but it also helps to minimize the tilt energy of the twisted 
chiral rods, which are otherwise frustrated on a flat membrane.

In Ref\cite{powers1}, it was shown that a disk shaped colloidal membrane 
can transform into a twisted ribbon under external stretching force.  
Here the contribution of the chiral nematic field, spread over the ribbon shaped membrane, was approximated to the energy of the open edges of the ribbon, assuming an uniform orientation of the rods in the bulk of the ribbon. It was argued that in the limit of small twist penetration depth 
($\lambda_p$) all the twist is concentrated within a thin strip at the membrane edge. This could be a good approximation for disc shaped membranes \cite{measured09,Gibaud,meyer} of typical radius $10\mu m$ and penetration depth $~\sim 0.5\mu m$, which indeed shows chiral twist near the periphery and nearly uniform orientation of the rods in the bulk. However ribbons
have typical half width of $\sim 1\mu m$ \cite{Gibaud}, comparable to 
$\lambda_p$, and also due to local membrane curvature the rods are not aligned aross the width  of the ribbon.
 
Here we take the reverse approach, i.e., assuming the membrane shape to 
be a minimal surface we seek the true lowest energy configuration of the 
chiral nematic field. We formulate a covariant theory for the 
chiral nematics on the curved surface and show, by minimizing 
the Frank free energy, that nontrivial nematic patterns can 
emerge on helicoid and catenoid shaped membranes.

The membrane surface, embedded in three dimensions, is described by the position 
vector $\vec R = [x(u,v), y(u,v), z(u,v)]$, where the two independnet parameters $(u,v)$ span the surface. The vectors defining the local 
tangent plane on the surface are
 $\vec u = \frac{\partial \vec R}{\partial u} = (x_u,y_u,z_u),$ and 
$\;\; \vec v = \frac{\partial \vec R}{\partial v} = (x_v,y_v,z_v)$
where subscripts denote partial derivatives,
The metric $g_{ij}$ on the this curved surface is obtained from the line 
element, $d s^2  = dx^2 + dy^2 + dz^2$. Expanding the differentials
as functions of the parameters $u,v$ yields, 
\begin{eqnarray}
    d s^2  &=& 
   \Big{(}x_u du + x_v dv \Big{)}^2 + \Big{(} y_u du + y_v dv \Big{)}^2 + \Big{(}z_u du + z_v dv\Big{)}^2   \nonumber \\ 
     &=&  (x_u ^2 + y_u^2 + z_u^2  ) du^2 +  (x_v ^2 + y_v^2 + z_v^2) 
dv^2   \nonumber\\ 
 &&\;\;\;\;\;\;\;\;\;\;\;\;\;\;\;\;\;\;\;\;\;\;\;\;\;\;\;\;\;\;+\;2(x_ux_v + 
y_u y_v + z_u z_v) du\; dv\nonumber\\
&\equiv& g_{ij} dx^i dx^j,
\end{eqnarray}
Here $g_{ij}$ ($i=1,2$) defines the $2\times2$ symmetric metric tensor on the surface, where $dx^1=du$ and $dx^2=dv$. The coefficients of $du^2,dv^2$ and $dudv$ (above) are identified as $g_{uu}, g_{vv}, 2g_{uv} 
(= 2g_{vu})$, respectively. If $\vec u$ and  $\vec v$ are orthogonal,
 and are along the principal axes, which is the case for all our applications, then $g_{ij}$ is diagonal.
It will be convenient to work with the  normalised unit vectors 
$\hat u=\vec u/|\vec u|,\;\hat v=\vec v/|\vec v|$, and in our case
$|\vec u|=\sqrt{g_{uu}}$ and $|\vec v|=\sqrt{g_{vv}}$. Although our 
nematic vector field $\hat m(u,v)$ is defined on the surface, both 
$\hat m(u,v)$ as well as its covariant tensorial derivatives (e.g., 
$\vec{\nabla} \times \hat m$) will have non-zero components 
perpendicular to the surface. Therefore we define unit normal
to the surface $\hat{w} = \hat{u} \times \hat{v}$. The resulting 
local orthogonal frame ($\hat u,\hat v,\hat w$) can now be used 
for describing $\hat m(u,v)$ and its covariant derivatives. 
Since $\hat w$ is the surface normal our curved surface can be 
defined as $w= w_0$ (constant) and therefore derivatives along the 
normal $\partial /\partial w$ yield zero. 
We will now generalise the line element $ds^2$, mentioned above, 
for the 3D space by adding $dw^2$ to it. 
Thus $g_{ij}$ is now enlarged to a $3\times 3$ matrix with 
$g_{ww}=1$, and the corresponding non-diagonal components 
$g_{uw}=g_{vw}=g_{wu}=g_{wv}=0$.
 
\subsection{Frank free energy on curved surface} 
Frank free energy of the nematic field $\hat m(u,v)$ on a curved surface
can be written as 
\begin{eqnarray}
&F&= \int\int\sqrt{g} \: du \: dv  \frac{1}{2}\Big{[}
K_1 (\vec{\nabla}. \hat{m})^2 + K_3 (\hat{m} \times \vec{\nabla} \times \hat{m} )^2\nonumber\\
&&    +  K_2 (\hat{m}.\vec{\nabla} \times \hat{m} -q)^2 
- K_{24}\hat\nabla .[\hat m (\hat\nabla.\hat m )+\hat{m} \times \vec{\nabla} \times \hat{m}]\nonumber\\    
 &&   - C (\hat{m}.\hat{w})^2   \Big{]}\;, 
    \label{eq.F}
\end{eqnarray}

where $\sqrt{g}dudv$ is the invariant area element ($g$ being the determinant of $g_{ij}$).  $K_1,K_2$ and $K_3$
are the splay, bend and twist modulus and henceforth we will work in the standard one constant approximation  $K_1=K_2=K_3=K$. Further, $q$ is the intrinsic chirality of the nematic directors and $C$ is the strength of 
the depletion interaction which promotes nematic allignment along the surface normal $\hat w$. Note that the chiral term which is linear in $q$ will
couple membrane curvature with nematic chirality \cite{selingerPRL,HFterm} via the use of covariant definition of curl. Since we work with finite surfaces with bounds $u\in [u_i,u_f]$ and $v\in [v_i,v_f]$, we also
include the saddle-splay boundary term with modulus $K_{24}$ The term $\hat{m} \times \vec{\nabla} \times \hat{m}$ can 
also be written as $-(\hat m.\hat\nabla)\hat m$. Particularly, when the nematics prefers to orient normal to the surface (which is promoted
by the depletion interaction here), the $K_{24}$ has been shown to be important \cite{k24-90,k24-91}. Since  all the modulii have the same dimension we will express them in units of $K$. Henceforth 
we use $K\equiv 1$ and $K_{24}/K\equiv K_{24}$.

The expressions for the covariant divergence and curl on
curved surfaces are \cite{softgij},  
\begin{eqnarray}
    \textbf{Div}\; (\hat {m})&=&\vec{\nabla} . \hat {m} = \frac{1}{\sqrt{g}} \partial_i  \bigg{(}       \sqrt{\frac{g}{g_{ii}}}  {m}^i \bigg{)}.\nonumber\\
 \textbf{Curl}\; (\hat m)&=&\vec{\nabla} \times \hat {m} = \sqrt{\frac{g_{ii}}{g} }    \: \epsilon^{i j k} \:  \partial_j  \bigg{(} \sqrt{g_{kk}}  {m}^k \bigg{)}\: \hat{e}_i.
 \label{Eq.divcurl}
\end{eqnarray}
Here $\{m^i\}$ are the components of the nematic vector expressed in terms of the normalised unit vectors $\hat{m} = m_u \hat{u} + m_v \hat{v} + m_w \hat{w}$. $\epsilon ^{ijk}$ are the contravariant Levi-Civita-symbols, 
$\{\hat e_i\}$ are the unit vectors. In both the expressions above, sum is implied over repeated indices, however $g_{ii}$ denote diagonal metric elements.  The indices $(1,2,3)$ stand for the triad ($\hat{u},\hat{v},\hat{w})$. 

\subsection{Flat Disk}
We first consider the simpler case of a circular, flat membrane as a
disc of radius $R$. 
In cylindrical polar coordinates $(u,v,w)\equiv (r,\phi,z)$ and the surface is given by $\vec R=(r\cos\phi, r\sin\phi, 0)$,
with $r<R$. The corresponding unit vectors $(\hat u,\hat v,\hat w)$ are
the standard polar $(\hat r,\hat\phi,\hat z)$,  
and the diagonal metric tensor comes from the line element $d s^2 = dr^2 + r^2 d\phi^2 + dz^2$. While defining the orientation of the nematic director field all previous theoretical works \cite{meyer,lubenskyRaft} had assumed the director $\hat m(r)$ to lie in the $(\hat\phi,\hat z)$ plane making 
an inclination angle $\phi(r)$ with $\hat z$. In addition azimuthal symmetry (i.e., no dependence on $\phi$) was also assumed. Here we relax the first
assumption and allow $\hat m$ to have a component along $\hat r$ making an angle $\alpha(r)$ with the ($\hat\phi,\hat z$) plane. See Fig.\ref{fig:Disk2}-a. 
Thus the director field is 
\begin{equation}
\hat{m}[\phi(r),\alpha(r)] = (\sin\alpha,\cos\alpha\sin\phi,\cos\alpha \cos\phi).
\end{equation}
Note that due to nematic symmetry i.e., invariance under inversion
$\hat m\rightarrow -\hat m$, the orientation $(\phi,\alpha)$ is same 
as $(\phi\pm \pi,-\alpha)$. Therefore, the allowed ranges for these angles
can be $\phi\in [0,\pi]$ and $\alpha\in [-\pi/2,\pi/2]$. However in some
plots later we allow the range $\phi\in [-\pi,\pi]$ so that the variation
of nematic orientation appears continuous when $\phi$ goes across zero.
Mapping the negative half of $\phi$ to the positive half by the shift
$(\phi\rightarrow \phi+\pi,\alpha\rightarrow -\alpha)$ makes $\phi$ 
variation appear discontinuous and $\alpha$ suffers a slope discontinuity 
at the origin $(\phi=0,\alpha=0)$, although spatially the variation is smooth. Using this parameterization for $\hat m(r)$ and the metric 
elements of the surface, Eq.\ref{Eq.divcurl} yields
\begin{eqnarray}
\vec{\nabla}.\hat {m} &=& \alpha '(r) \cos \alpha + \frac{\sin \alpha}{r}\nonumber\\
  \vec{\nabla} \times \hat {m}&=& \Big{\{}\frac{\cos \alpha}{r}(r \phi ' \cos \phi + \sin \phi )-\alpha ' \sin \alpha \sin \phi,\nonumber\\ 
  && \alpha ' \sin \alpha \cos \phi + \phi '\cos \alpha \sin \phi,\;0 \Big{\}}
\end{eqnarray}
The resulting Frank free-energy for ths system is then, 
\begin{eqnarray}
&F&= 
\int dr \frac{\pi}{r}\Big\{K_1 \big{[}r \alpha ' \cos \alpha +\sin \alpha\big ]^2 \nonumber\\
 &&+ K_2 \big [q r-\cos ^2 \alpha (r \phi '+\sin \phi \cos \phi )
 \big ]^2  \nonumber\\  
 &&+ \frac{K_3}{4} [r \{-4 \alpha ' \sin 2 \alpha \sin ^2\phi +4 r 
 \alpha '^2 \sin ^2\alpha  + \sin ^2 2 \alpha \phi ' \nonumber\\
 &&\times ( r \phi '+\sin 2 \phi) \}  
 + \sin ^2\phi  \left(4 \cos ^4\alpha \sin ^2 \phi+\sin ^2 2 \alpha 
\right)]  \nonumber\\ 
&&-K_{24} r [\frac{1}{2} \alpha ' \sin 2 \alpha (\cos 2 \phi-5)-2 r \alpha '^2 \cos 2 \alpha \nonumber\\
&& -r \alpha '' \sin 2 \alpha + \cos ^2\alpha \phi ' \sin 2 \phi] 
- C r^2 \cos ^2\alpha  \cos ^2 \phi     
 \Big\}\nonumber\\
\end{eqnarray}
Extremization of this free energy $F=\int dr \cal F$ (where $\cal F$ is the 
free energy density) leads to two Euler-Langrange equations
which are  $\frac{\partial}{\partial r}
[\frac{\partial \cal F}{\partial \phi'(r)}]-\frac{\partial \cal F}{\partial \phi} = 0$ and $ \frac{\partial}{\partial r}
[\frac{\partial \cal F}{\partial \alpha'(r)}]-\frac{\partial \cal F}{\partial \alpha} =0$. Nondimensionalised forms of these equations 
for the $\phi$ and $\alpha$ fields, respectively, are  
\begin{figure}
\centering
\includegraphics[width=0.5\textwidth, angle=0]{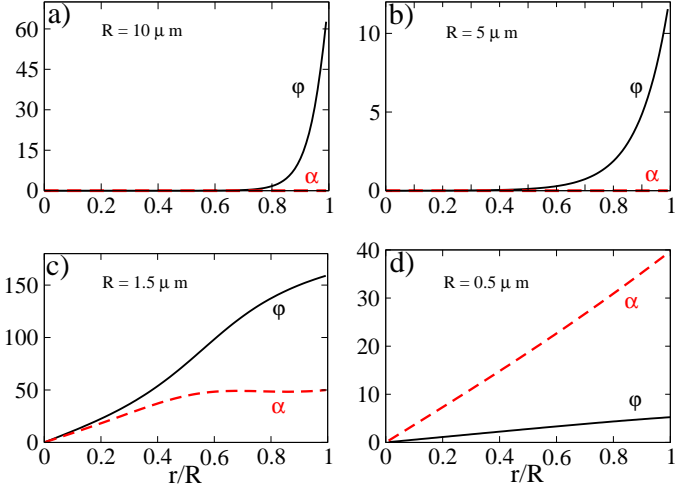}
\caption{   
Inclination angles $\phi (r)$ and 
$\alpha (r)$ (in degrees) as functions of the scaled radial 
distance $r/R$ from the center of the disk. (a-d) show solutions
for different radii (see legends). Nonzero $\alpha$ is found
only for small radii, much smaller than experimentally obtained
disk radii $\sim 10\mu m$. System properties, the penetration 
depth $\lambda_p=0.5\mu m, q=0.5 \mu m ^{-1}$ and   
$K_{24}=1$ are held fixed. Two typical nematic configurations, 
for $R=10\mu m$ and $1.5\mu m$, are shown in fig.\ref{fig:Disk2}.}
\label{fig:Disk1}
\end{figure}  
\begin{figure}
\centering
\includegraphics[width=0.45\textwidth, angle=0]{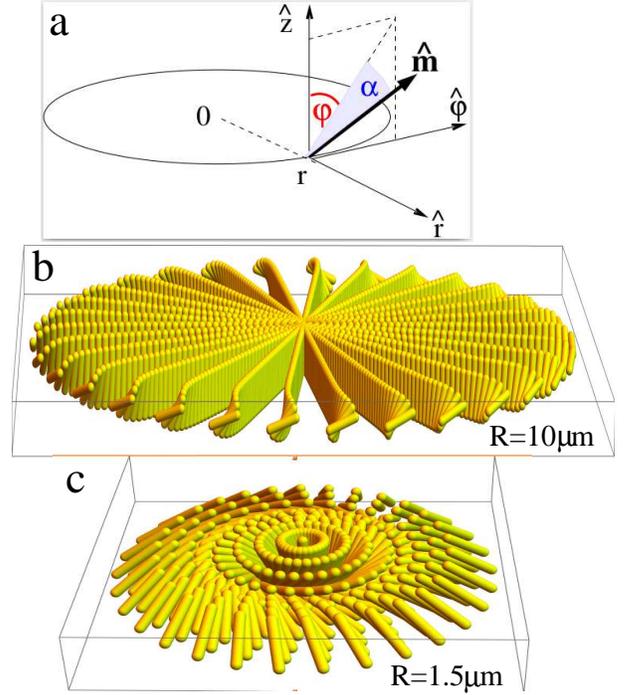}
\caption
{Nematic configurations for different disk radii at fixed
$\lambda_p=0.5\mu m,q=0.5 \mu m ^{-1}$ and  $K_{24}=1$. a) defines the 
angles $\phi$ and $\alpha$, (b-c) are for disk radii $R=10$ and 
$1.5\mu m$. b) shows the deviation from vertical only at the 
periphery as seen in PolScope measurements \cite{measured09}. 
c) shows exotic nematic orientation where the directors undergo 
large deviation from the normal. As we move radially outwad, 
the director first turns parallel to the $x-y$ plane and then 
rises up, mainitaining clockwise rotation all through, due to 
its chirality.} 
\label{fig:Disk2}
\end{figure}
\begin{eqnarray}
&&\frac{1}{r}\{\cos ^2 \alpha [\sin \phi \{2 q r \sin \phi+(r^2+1) 
\cos \phi\}
-r^2 \phi ''(r)\nonumber\\ &&+ r \phi '(r)]\}-r \alpha '(r)\left(q-\phi '(r)\right) \sin 2 \alpha =0,\;\nonumber\\ 
&&\;\;\;\;\;\;\;\;\;\;\;\;\;\;\;\;\;\mbox{and} \nonumber\\
&&\frac{1}{4 r}\{4 q r^2 \sin 2 \alpha \phi '(r)+2 q r \sin 2 \alpha \sin 2\phi-4 r^2 \alpha ''(r) \nonumber\\
&&-2 r^2 \sin 2 \alpha \phi '(r)^2+r^2 \sin 2 \alpha \cos 2\phi+r^2 \sin 2 \alpha \nonumber\\ &&-4 r \alpha '(r)+
\sin 2 \alpha \cos 2\phi+\sin 2 \alpha\}=0.
\label{eq.EL}
\end{eqnarray}
Here the radius and the intrinsic chirality have been rescaled as $r'=r/{\lambda_p}$ and $q' =q\lambda_p$, by the twist penetration depth 
$\lambda_p=\sqrt{\frac{K}{C}}$ \cite{meyer}. However, in what follows, 
we have omitted the
primes to keep the notations simple. Note that the saddle-splay term, involving $K_{24}$, being a boundary term does not affect these equations
but will show up in the boundary conditions (BC). 
The BC are as follows. The directors are assumed to be normal to the 
disc at its center i.e., $\phi(r=0)=\alpha(r=0)=0$ and torque free BC 
\cite{kaplan2010theory} are imposed at the periphery, i.e.,
$\frac{\partial \cal F}{\partial \phi'(r)}=0$ and $\frac{\partial \cal F}{\partial \alpha'(r)}=0$, at $r=R$. In order to avoid numerical problems 
at $r=0$, we use $r=r_0=10^{-6}$. 
The torque free BC at the periphery amounts to  
\begin{eqnarray}
\cos ^2\alpha (R) \big[q R - \frac{K_{24}+1}{2} \sin 2\phi (R) -
R \phi '(R)\big] =0 \;,\label{bc1-disk}\\
\mbox{and}\;,\; \frac{\sin 2 \alpha (R) }{4} \big[1-5 K_{24} + (1+K_{24}) \cos 2 \phi (R)\big] \nonumber\\
 + R \alpha '(R) \left [1-2 K_{24} \cos 2 \alpha(R)\right]  =0 \label{bc2-disk}
\end{eqnarray}

In the limit $\alpha=0$ and $K_{24}=0$, the first of these BCs' reduces to a form which is equivalent to $\hat{m}.\vec{\nabla} \times \hat{m} -q=0$, and the second one gives null.

We obtain numerical solution of this nonlinear, boundary value problem (Eq.\ref{eq.EL} along with Eq.\ref{bc1-disk},\ref{bc2-disk}) using Mathematica.
The angular orientation fields for flat circular discs have been measured experimentally using retardance \cite{measured09}. It was found that the directors are nearly normal in the bulk of the disk and show large tilt, along the $\hat\phi$ direction, near the disk periphery. 
Our numerical solutions for the $R/\lambda_p>>1$ case shows the same exponential rise in inclination $\phi$ within the thin twist penetration layer, see Fig.\ref{fig:Disk1}a,b and the corresponding director configuration in Fig.\ref{fig:Disk2}b.
The inclination $\alpha(r)$ however turns out to be zero for these
disks, unless the disk radius is unrealistically small, as shown in  
Fig.\ref{fig:Disk1}c,d. The corresponding director configuration is
shown in  Fig.\ref{fig:Disk2}c, which is indeed exotic but probably
is difficult to realise in experiments.

\subsection{Helicoid}
\begin{figure}
\centering
\includegraphics[width=0.45\textwidth, angle=0]{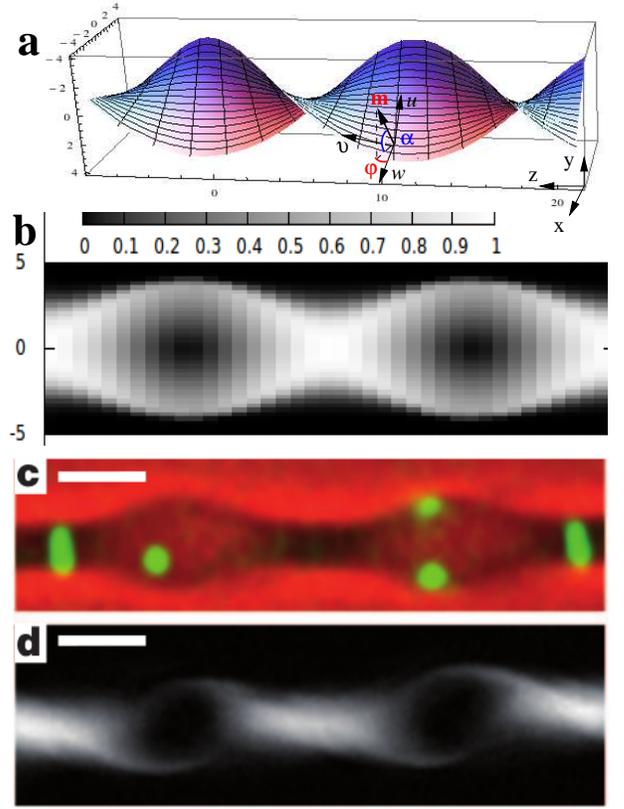}
\caption{\label{fig:heliaxix}{(a) shows local $(\hat u,\hat v, \hat w)$ 
axes and the angles $\phi,\alpha$ on the helicoidal surface. (b) is 
our theoretical plot corresponding to the $\phi$ profile in our 
Fig.\ref{fig:Cube}-a. (c,d) are experimental pictures from 
Ref\cite{Gibaud} (with permission). c) shows fluorescence (green) 
image of few rods, nearly parallel and perpendicular to
the image plane at the helicoid neck and the bulge, respectively. 
d) shows LC-PolScope image of the rods where the intensity at each 
pixel is proportional to $\sin ^2\theta$, where $\theta$ is the local 
tilt angle between the rod and the normal to the image plane 
(scale bars $2\mu m$). The small up-down asymmetry in (d) is
because LC-PolScope image captures signal from one plane. Typically 
a stack of images are collected to infer the 3D structure
(see for example, Fig.4e or f in Ref\cite{Gibaud}). 
In (b), we show the same quantity, $\sin ^2\theta$ (color bar [0,1]), integrated over the 3D helicoid, 
corresponding to our solution in Fig.\ref{fig:Cube}-a, at fixed 
penetration depth $\lambda_p=0.5\mu m, q=0.5 \mu m^{-1}$ and $K_{24}=1$.
Only small, gradual, deviation from normality occurs towards the
edge of the helicoid. Here we treated $y-z$ plane as the image 
plane, with its normal along $\hat x$. 
}}
\end{figure}

Now we discuss the director arrangement on a helicoidal membrane surface
which is also known as a twisted ribbon.
The parametric equation for the helicoidal surface, with pitch $2\pi C_1$,
in the cartesian frame, is $\vec R =( u \cos v,\; u \sin v,\; C_1 \ v)$,
where $v$ is the angular coordinate and $u\in [-u_f,u_f]$ spans the width 
of the helicoid.
The unit vectors $\hat u$ and $\hat v$, constituting the tangent plane,
and the surface normal $\hat{w} = \hat{u} \times \hat{v}$ are, 
\begin{eqnarray}
&&\Big {(} \cos{v}, \sin{v}, 0 \Big{)} \;,
\frac{1}{\sqrt{u^2+c^2}} \Big {(} -u \sin{v}, u \cos{v}, C_1 \Big {)}\;, 
\nonumber\\
&& \mbox{ and }\frac{1}{\sqrt{u^2+C_1^2}}  \Big {(} C_1 \sin{v}, -C_1 \cos{v}, u \Big {)}. \label{eq.huvw}
\end{eqnarray}
The triad $(\hat u,\hat v, \hat w)$ is shown in Fig.\ref{fig:heliaxix}a
on the helicoid surface. From the line element on the surface $d s^2 = d u^2 + (u^2 + C_1^2) d v^2 + d w^2$ we identify the components of the diagonal metric tensor. The director field in this $(\hat u,\hat v, \hat w)$ frame using $\phi$ and $\alpha$, defined as before in the disc case, is
\begin{equation}
\hat{m}[\phi(u),\alpha(u)] = (\sin\alpha,\cos\alpha\sin\phi,\cos\alpha \cos\phi). \label{eq.hm}
\end{equation}
Here, azimuthal symmetry i.e., dependence only on $u$, is assumed. As before, the projection of the director on $(\hat v,\hat w)$ plane makes angle $\phi$ with $\hat w$ (surface normal) and $\alpha$ is the 
inclination of the director to the $(\hat v,\hat w)$ plane, see 
Fig.\ref{fig:heliaxix}-a. Thus the director has a
component $\sin\alpha$ along $\hat u$. Using this parameterization for $\hat m(u)$ and the metric elements of the surface, Eq.\ref{Eq.divcurl} yields
\begin{eqnarray}
\vec{\nabla} . \vec {m} &=& \frac{\partial_u (\sqrt{u^2 + C_1^2}  m_u)}{\sqrt{u^2 + C_1^2}} + \frac{\partial_v m_v}{\sqrt{u^2 + C_1^2}} + \partial_w  m_w\nonumber\\
 \vec{\nabla} \times \vec {m} &=& ( \frac{1}{\sqrt{u^2 + C_1^2}}\; \partial_v m_w - \partial_w m_v\;, \;   \partial_w m_u - \partial_u  m_w \;, \nonumber\\
 &&  \frac{1}{\sqrt{u^2 + C_1^2}}  (\partial_u (\sqrt{u^2+C_1^2} m_v) - \partial_v m_u ) 
\end{eqnarray}
Here $\vec{m} = m_u \hat{u} + m_v \hat{v} + m_w \hat{w}$ and 
$\partial_w \vec{m} =0$. 

\begin{figure}
\includegraphics[width=0.45\textwidth, angle=0]{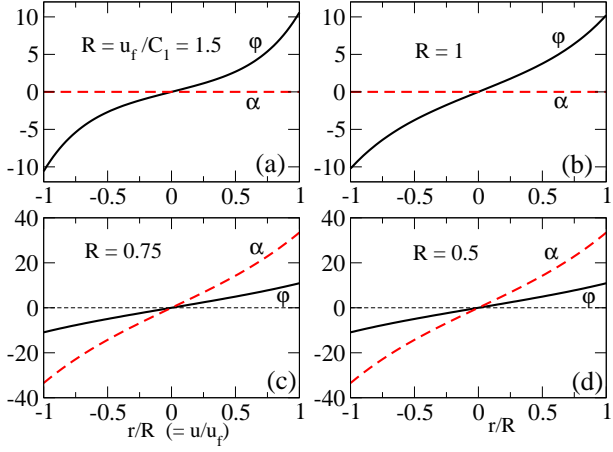}
\caption{
$\phi (r)$ and $\alpha (r)$ (in degrees) as functions of the scaled 
"radial" distance $u/u_f$ (along the curved $u$ axis), at fixed 
$\lambda_p=0.5\mu m, q=0.5 \mu m ^{-1}$ and $K_{24}=1$. 
(a-d) show solutions
for different helicoid widths $u_f/C_1$ (see legends), $2\pi C_1$
being the pitch. Nonzero $\alpha$ is predicted only for narrow 
helicoids (c and d) with small $u_f/C_1$.  Experimentally obtained 
ones \cite{Gibaud} have 
$u_f/C_1\sim 1$. Two representative nematic arrangements on
helicoids are shown in Fig.\ref{fig:Cube5}. }
\label{fig:Cube}
\end{figure}

\begin{figure}
\centering
\includegraphics[width=0.4\textwidth, angle=0]{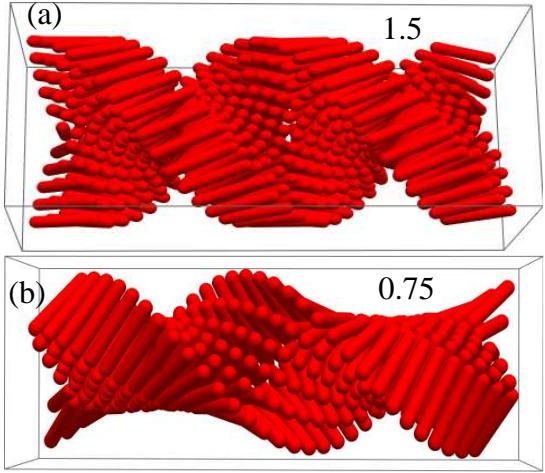}
\caption{\label{fig:Cube5}{(a,b) Shows nematic arrangements on 
helicoidal surfaces, corresponding to Fig.\ref{fig:Cube}-a and -c. 
The scaled helicoidal widths $u_f/C_1$ are mentioned in the 
legends. In (b) deviation from the  $(v,w)$ plane (i.e., nonzero
$\alpha$) is manifest. }} 
\end{figure}

Following the same recipe as before, i.e., defining 
$\lambda_p = \sqrt{\frac{K}{C}}$, using dimensionless variables 
$r \equiv u/C_1,\; q\equiv q\lambda_p$, and 
$C_1\equiv C_1/\lambda_p$  the Frank free energy density is
\begin{eqnarray}
&&{\cal F}=\frac{1}{2 \left(r^2+1\right)^{3/2}}\Big\{K_1 [\left(r^2+1\right) \alpha ' \cos \alpha + r \sin \alpha]^2\nonumber\\
&&+ K_2 [C_1 q \left(r^2+1\right)-\cos ^2 \alpha \left(\left(r^2+1\right) \phi ' + r \sin \phi \cos \phi \right) ]^2 \nonumber\\
&&+ K_3 [(r^2+1) ((r^2+1) \alpha '(r)^2 \sin ^2\alpha 
+ \sin ^2\alpha \cos ^2 \alpha \phi ' \nonumber\\ 
&&\;\;\;\;\;\;\;\;\;\;\times((r^2+1) \phi '+r \sin 2 \phi )
- r \alpha ' \sin 2 \alpha  \sin ^2\phi ) \nonumber\\
&&\;\;\;\;\;\;\;\;\;\;+ r^2 \cos ^4\alpha \sin ^2\phi (\tan ^2 \alpha +\sin ^2\phi )] \nonumber\\
&& -K_{24} [ (\sin ^2 \alpha - 
         \cos ^2 \alpha \sin ^2\phi) + \frac{(1 +  r^2)}{2}\nonumber\\
&& \;\;\;\;\;\;\;\;\;\;\times\{r (5 - \cos 2\phi) \sin 2\alpha \alpha ' + 
         4 (1 + r^2) \cos 2\alpha \alpha '^2 \nonumber\\
&& \;\;\;\;\;\;\;\;\;\;-2 r \cos ^2 \alpha \sin 2\phi \phi ' + 
         2 (1 + r^2) \sin 2\alpha \alpha ''\}]\nonumber\\
&&- C C_1^2 \left(r^2+1\right)^2 \cos ^2 \alpha \cos ^2\phi\Big\}.
\end{eqnarray}
This free energy density is expected to be symmetric with respect to 
$r\rightarrow -r$, i.e., the two strips of the helicoid at positive and negative $r$ should have the same free energy density. Mathematically, 
this can be realised if  $\phi,\alpha$ are odd functions of $r$, and consequently $\phi ',\alpha '$ are even functions. Later we show that 
free energy minimization indeed leads to such solutions for both helicoids
and catenoids (to be discussed later).  Using one constant approximation, the Euler-Lagrange's equations yield,
\begin{eqnarray}
&\frac{1}{(r^2+1)^{3/2}}\{\cos \alpha \{\cos \alpha (\sin \phi (({C_1}^2 (r^2+1)^2+r^2-1) \cos \phi \nonumber\\
&+ 2 {C_1} qr (r^2+1) \sin \phi)-r(r^2+1) \phi '(r)-(r^2+1)^2 \phi ''(r))
\nonumber\\
&-2 (r^2+1)^2 \alpha '(r) \sin \alpha ({C_{1}} q-\phi '(r))\}\} = 0. 
\end{eqnarray}
and,
\begin{eqnarray}
&\frac{1}{\sqrt{r^2+1}}\{\sin 2 \alpha (({C_1}^2 (r^2+1)^2+r^2-1) \cos ^2\phi+\nonumber\\
&{C_1} q r(r^2+1) \sin 2\phi+(r^2+1)^2 \phi '(r) (2 {C_1} q-\phi '(r)))-\nonumber\\
&2 (r^2+1) ((r^2+1) \alpha ''(r)+r \alpha '(r))\}= 0.
\end{eqnarray}
The BC are similar as before, i.e., at the central line  
of the helicoid $\alpha (r=0)=\phi (r=0)=0$. As before, the open
edges of the helicoid, at $r=\pm R=\pm u_f/C_1$, are considered 
to be torque free, which yields, 
\begin{eqnarray}
&& \cos ^2\alpha \left[\phi ' - C_1 q +
(K_{24}+1) \frac{R\sin 2\phi}{2(R^2+1) } \right] = 0,\nonumber\\
&&\;\;\;\;\;\;\;\;\;\;\;\;\;\; \;\;\;\;\;\;\;\;\;\;\;\;\;\;\;\; 
\mbox {and,}\nonumber\\
&&\frac{R\sin 2 \alpha} {4(R^2+1)}\Big[1 -5 K_{24}+ (1+K_{24}) \cos 2 \phi\Big]\nonumber\\
&& \;\;\;\;\;\;\;\;\;\;\;\;\;\; \;\;\;\;\;\;\;\;\;\;\;\;\;\;\;\;
+ \alpha ' [1-2 K_{24} \cos 2 \alpha ]  = 0\;.
\end{eqnarray}
Due to spatially varying principle curvatures (although the mean 
curvature is zero), the solutions for helicoids, obtained here, are
 more complex. Qualitatively two different type of solutions emerge : 
 1) with $\alpha=0$  (shown in Fig.\ref{fig:Cube}-a,b), and 
 2) $\alpha\neq 0$ (shown in Fig.\ref{fig:Cube}-c,d), while $\phi$ 
remains small for both cases. Two representative nematic configurations,
corresponding to Fig.\ref{fig:Cube}-a and c are shown in 
Fig.\ref{fig:Cube5}. Experimentally obtained helicoids (twisted ribbons)
\cite{Gibaud} have pitch $\sim 6\mu m =(2\pi C_1)$ and width 
$\sim 2\mu m = (2 u_f$), which yields $u_f/C_1\sim 1$. 
We therefore pick  Fig.\ref{fig:Cube}-b for computing pixel 
intensities $I\propto\sin^2 \theta$, as detected by LC-PolScope images,
where $\theta$ is the angle between the nematic director and the normal
($\hat x$) to the image plane ($y-z$) here. For example, the part of
the ribbon where the directors point along $\hat x$ should appear
dark. The computed intensity
map is shown in Fig.\ref{fig:heliaxix}-b and compared with 
experimental image \cite{Gibaud} in Fig.\ref{fig:heliaxix}-d. 
The intensity is obtained by computing $\cos\theta = \hat m .\hat x$,
where $\hat x$ pertains to the cartesian lab frame. Using 
Eq.\ref{eq.huvw} and \ref{eq.hm} we get,
\begin{eqnarray}
\cos\theta &=&\sin\alpha\cos v+ 
\frac{\cos\alpha\sin v}{\sqrt{1+(u/C_1)^2}}
\{\cos\phi - \frac{u}{C_1}\sin\phi \}, \nonumber\\
&&\mbox{ which for $\alpha=0$ yields,}\nonumber\\
\sin ^2\theta &=&1 - \frac{\sin ^2 v}{1+(u/C_1)^2}
(\cos\phi - \frac{u}{C_1}\sin\phi )^2.
\end{eqnarray}
Note that since $\phi$ and $\alpha$ are odd functions of $u$,
under the  change $u\rightarrow -u$ the intensity remains
same. It implies that the strips belonging to positive 
and negative $u$ parts of the ribbon must produce the 
same intensity. This is manifest in the theoretical map
Fig.\ref{fig:heliaxix}-b. However, in the experimental image 
(Fig.\ref{fig:heliaxix}-d) the asymmetry (as explained in 
the caption) occurs due to the fact that a LC-PolScope 
image captures signal from a particular plane. Typically, 
by varying this plane a stack of images ($z$-stack) are 
collected to infer the 3D structure (such a stack is 
shown in Fig.4e and f in Ref-\cite{Gibaud}).


\section{Catenoid}
Now we consider the chiral nematic field on a catenoid shaped 
membrane surface.
Taking advantage of the azimuthal symmetry of the catenoid, the 
director field at different x-y planes can be measured using 
confocal microscopy and preliminary results indicate nontrivial
patterns \cite{apsabstract} with nonzero $\alpha$ in the bulk and
$\alpha\rightarrow 0$ at the free edges. 

The parametric equation for such a surface in cartesian frame is given by 
\begin{align}
\vec R= (C_1 \cosh{[\frac{v}{C_1}]} \cos{u}, C_1 \cosh{[\frac{v}{C_1}]}\sin{u}, \; v )\;,
\end{align}
\begin{figure}
\centering
\includegraphics[width=0.45 \textwidth, angle=0]{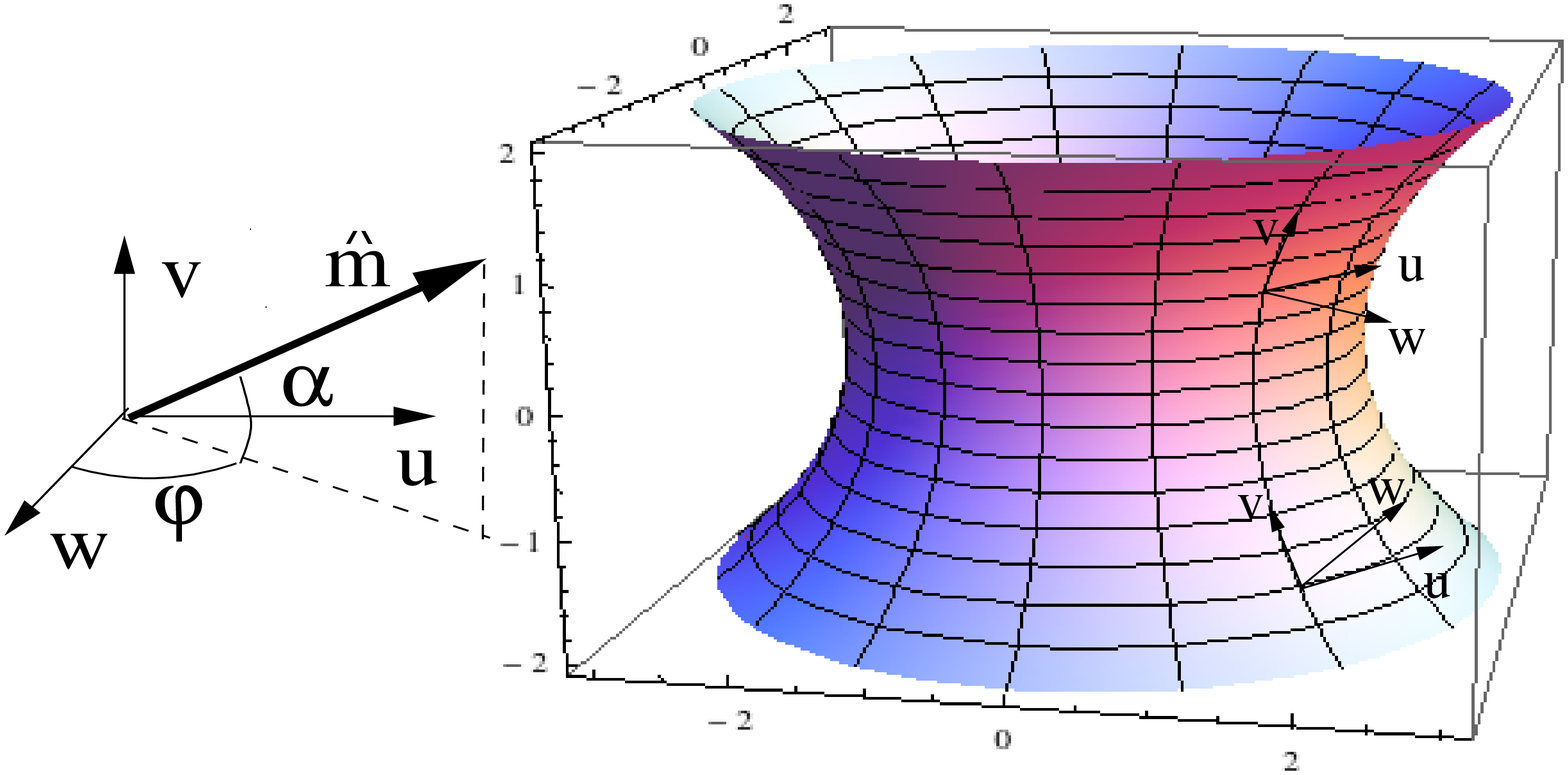}
\caption{{Local $(\hat u,\hat v, \hat w)$ axes on catenoid surface.}}
\label{fig:cataxis}\end{figure}
\begin{figure}
\centering
\includegraphics[width=0.45\textwidth, angle=0]{cat2bc.eps}
\caption{\label{fig:Cat1}{
$\phi (r)$ and $\alpha (r)$ (in degrees) as functions of the scaled 
distance $v/v_f$ (along curved surface). Here we have fixed 
$\lambda_p=0.5\mu m, q=0.5 \mu m ^{-1}$ and $K_{24}=1$, and 
applied two torque free BCs (see text). (a-c) show solutions
for catenoids of different aspect ratios $v_f/C_1$ (see legends), 
$C_1$ being the minimum neck radius. Experimentally observed ones 
\cite{apsabstract} have $v_f/C_1\sim 1.3$. With the two torque
free BCs (see text) we do not get $\alpha \neq 0$ solutions, 
contrary to experimental findings \cite{apsabstract}. Two representative nematic arrangements are shown in Fig.\ref{fig:Cat2}. (d) shows
a phase diagram with respect to two parameters $\lambda_p$ and 
$v_f/C_1$. Two type of solutions emerge, small $\phi$, as in 'a' 
or 'c' (open squares), and large $\phi$, as in 'b' (solid diamonds).  
}}
\end{figure}
\begin{figure}
\centering
\includegraphics[width=0.4\textwidth, angle=0]{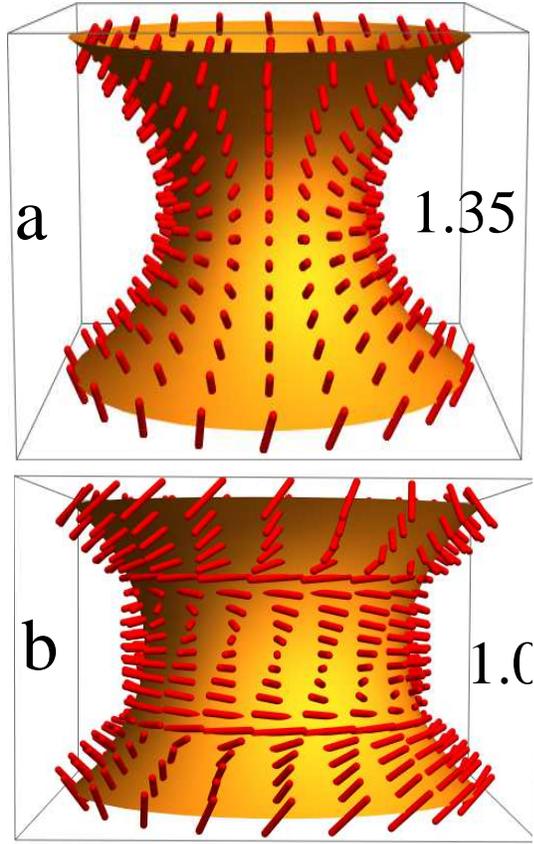}
\caption{\label{fig:Cat2}{ (a) and (b) show nematic arrangements (red)
on catenoid surfaces (yellow), corresponding to the small and large $\phi$ 
solutions, respectively, in Fig.\ref{fig:Cat1}. The aspect 
ratios $v_f/C_1$ are shown in the legends. In (b) the director 
rotates clockwise, as we move upwards along increasing $v$. }}

\end{figure}
\begin{figure}
\centering
\includegraphics[width=0.45\textwidth, angle=0]{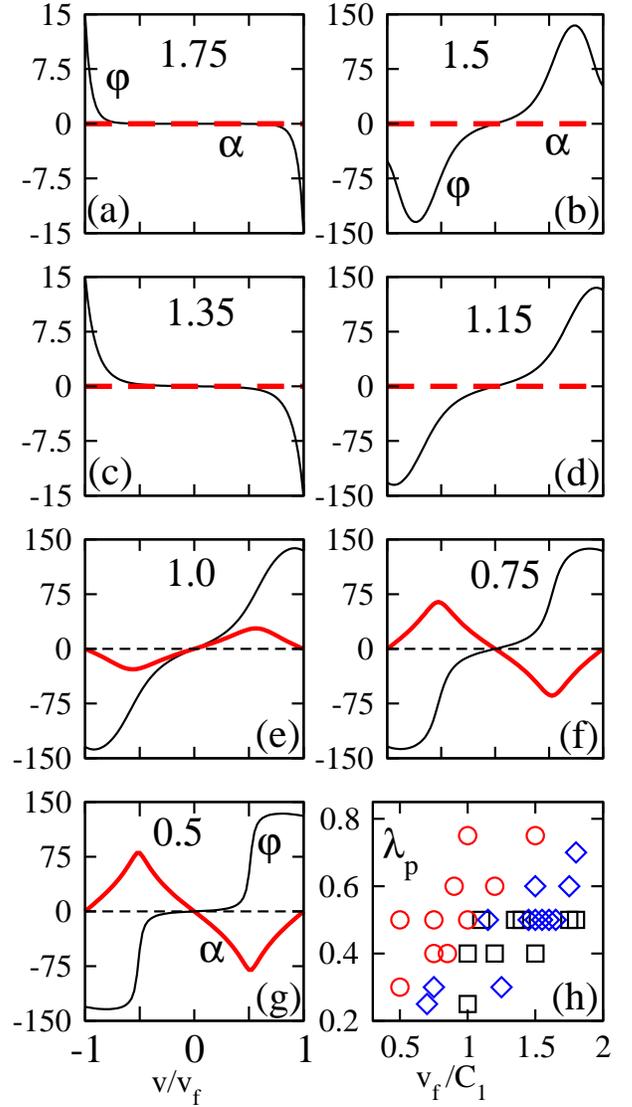}
\caption{\label{fig:Cat3}{$\phi (r)$ and $\alpha (r)$ (in degrees) as functions of the scaled distance $v/v_f$. We fixed 
$\lambda_p=0.5\mu m, q=0.5 \mu m ^{-1}$ and $K_{24}=1$, and employed 
torque free BC with respect to $\phi$, and set $\alpha (\pm v_f)=0$ at 
the open boundaries (edges). (a-g) are for different aspect ratios 
$v_f/C_1$ (see legends). 
Two representative nematic arrangements are shown in Fig.\ref{fig:Cat4}. 
(h) shows a phase diagram based on the two types of solutions that emerge 
in (a-g): 1) $\alpha\neq 0$ (red circles) at  $v_f/C_1\leq 1$,
and 2) $\alpha=0$ at larger $v_f/C_1\geq 1$. However, the $\alpha=0$ 
type has two sub-types, small $\phi$ (black squares) and large $\phi$ 
(blue diamonds), with no sharp boundary between them.}}
\end{figure}
\begin{figure}
\centering
\includegraphics[width=0.4\textwidth, angle=0]{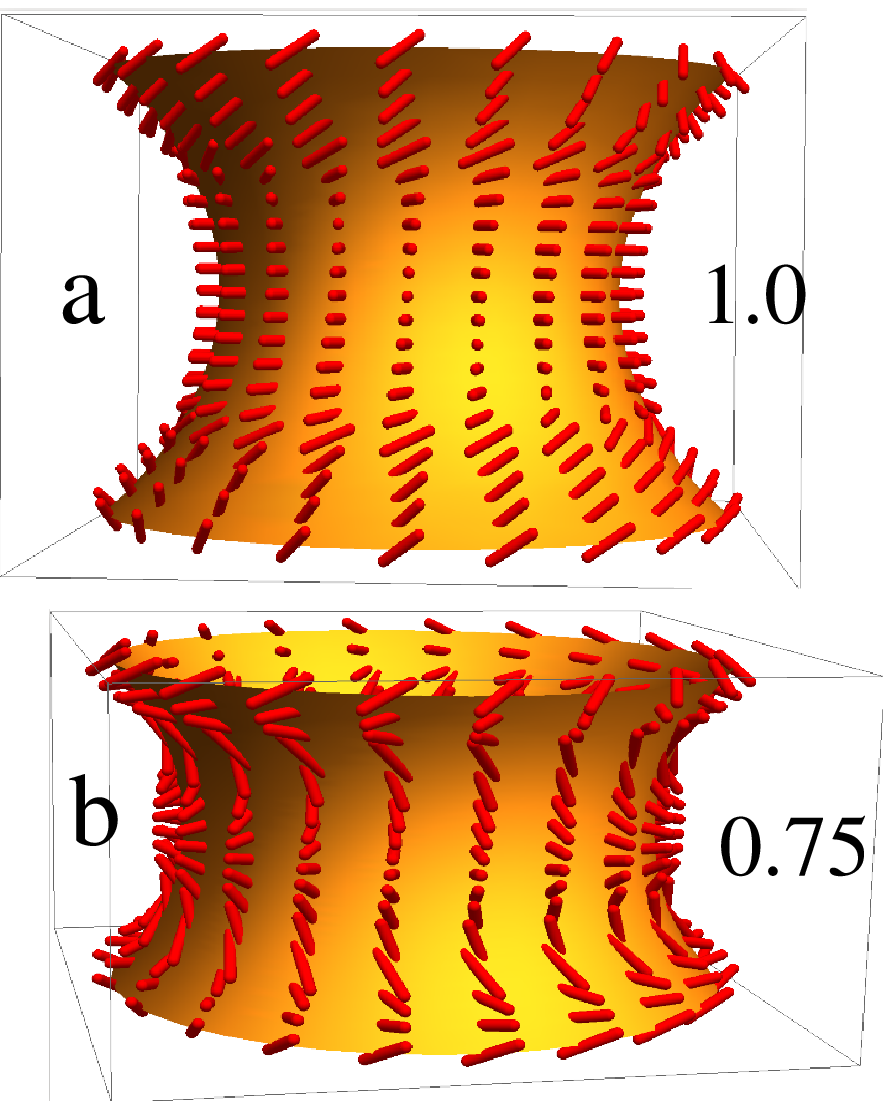}
\caption{\label{fig:Cat4}{(a) and (b) show nematic arrangements (red) 
on catenoid surfaces (yellow), corresponding to $\alpha (r) \neq 0$ 
cases in Fig.\ref{fig:Cat3}-e and f, respectively. The aspect ratios 
$v_f/C_1$ are shown in the legends. Unidirectional rotation of 
the director along increasing $v$ is visible.  }}
\end{figure}
where $u\in[0,2\pi]$ is the angular coordinate and $v\in [-v_f,v_f]$ 
is the distance along the longitudinal direction of the catenoid. 
$C_1$ is the cross sectional radius of the catenoid at $v=0$. 
The resulting triad $(\hat u,\hat v, \hat w)$, shown in 
Fig.\ref{fig:cataxis}, are
\begin{eqnarray}
\Big ( -\sin u, \cos u, 0 \Big ),\Big ( \text{tanh}{[\frac{v}{C_1}]} \cos u, 
  \text{tanh}{[\frac{v}{C_1}]}  \sin u, 
\text{sech}{[\frac{v}{C_1}]} \Big ),\nonumber\\ 
\mbox{and}\;\; \Big (\text{sech}{[\frac{v}{C_1}]} \cos u, 
\text{sech}{[\frac{v}{C_1}]}\sin u,   -\text{tanh}{(\frac{v}{C_1})} 
 \Big ). \nonumber
 \end{eqnarray}
Using line element, 
\begin{equation}
    d s^2 = \cosh^2 {(\frac{v}{C_1})} d v^2 + C_1^2 \cosh^2  {(\frac{v}{C_1})} d u^2 + d w^2
\end{equation}
we construct the metric. Further, the director field is defined using 
two angles $\phi$ and $\alpha$ (see Fig.\ref{fig:cataxis}), in this 
$(\hat u,\hat v, \hat w)$ frame, as
\begin{equation}
\hat{m}[\phi(v),\alpha(v)] = (\cos\alpha\sin\phi,\sin\alpha,\cos\alpha \cos\phi).
\end{equation}
Here, azimuthal symmetry i.e., no dependence on $u$, is assumed. Further,
the director does not lie fully in the $(\hat u,\hat w)$ plane, but may 
have a  nonzero component along $\hat v$.  The projection of the director 
on $(\hat u,\hat w)$ plane makes angle $\phi$ with $\hat w$ (surface normal) and $\alpha$ is the inclination 
of the director with the $(\hat u,\hat w)$ plane. Thus the director has a
component $\sin\alpha$ along $\hat v$. Using  $\hat m(v)$ and the metric elements of the surface, Eq.\ref{Eq.divcurl} yields,
after rescaling $v/C_1\rightarrow v$, 
\begin{eqnarray}
 \vec{\nabla} .\; \vec {m} &=& \frac{\text{sech}(v)}{C_1} [\alpha '(v) \cos \alpha +\tanh (v) \sin \alpha],\;\mbox{and} \nonumber\\
\vec{\nabla} \times \vec {m} &=& \Big\{\frac{\text{sech}(v)}{C_1}\{ \alpha '(v) \sin \alpha \sin \phi -\cos \alpha [\phi '(v) \cos \phi \nonumber\\
&+& \tanh (v) \sin \phi ]\},-\frac{\text{sech}(v)}{C_1}\{ \alpha '(v) \sin \alpha  \cos \phi \nonumber\\
&+&\cos \alpha \phi '(v) \sin \phi v \},0 \Big\}
\end{eqnarray}
Here $\vec{m} = m_u \hat{u} + m_v \hat{v} + m_w \hat{w}$ and 
$\partial_w \vec{m} =0$. 
Defining $\lambda_p = \sqrt{\frac{K}{C}}$ (where $K_1=K_2=K_3=K$) and 
using dimensionless variables $v\equiv v/C_1,\; q \equiv q\lambda_p$ and
$C_1\equiv C_1/\lambda_p$, 
the Frank Free energy reads, 
\begin{eqnarray}
&&{\cal F} = \frac{1}{2} \Big\{K_1 [\alpha ' \cos \alpha +\tanh v \sin \alpha ] ^2\nonumber\\
&&+ K_2 (C_1 q \cosh v+\cos ^2\alpha [\phi '+\tanh v \sin \phi \cos \phi ])^2 \nonumber\\
&&+K_3 (-\alpha '\tanh v \sin 2\alpha \sin ^2\phi +\alpha '^2 \sin ^2\alpha\nonumber\\ 
&&\;\;\;\;\;\;\;\;\;+\cos ^2\alpha (\phi '\sin ^2 \alpha   [\phi '+\tanh v \sin 2 \phi ] \nonumber\\
&&\;\;\;\;\;\;\;\;\;+\tanh ^2 v \sin ^2\phi (\cos ^2 \alpha  \sin ^2 \phi +\sin ^2\alpha )))\nonumber\\
&&-\frac{K_{24}}{C_1} \Big[\text{sech}\;v (\text{sech}^2 v \sin ^2 \alpha + 
        C_1 \cos ^2\alpha \;\text{sech}\;v \sin ^2\phi \nonumber\\
&&-    \sin ^2\alpha \;\text{tanh} ^2 v) + \alpha ' \tanh v\sin 2\alpha( \frac{\text{sech}\;v}{2} - C_1 \sin ^2\phi) \nonumber\\
&&  + \alpha '^2 \cos 2 \alpha (-C_1 + \text{sech}\; v) + C_1\phi ' \cos ^2\alpha \sin 2\phi \tanh v \nonumber\\
&&  + 
   \alpha ''  \cos \alpha\sin \alpha  (-C_1 + \text{sech}\;  v) \Big]\nonumber\\
&&\;\;\;\;\;\;\;\;\;-C C_1^2 \cosh ^2v \cos ^2\alpha  \cos ^2\phi \Big\}
\end{eqnarray}
Here we retained the different $K_j$s' to keep track of the splay, twist
and bend contributions. As mentioned before, the free energy density has the $v\rightarrow -v$ symmetry provided $\phi,\alpha$ are odd functions of $v$. 
The corresponding Euler-Langrange equations are,
\begin{eqnarray}
&&\frac{1}{2} \cos ^2 \alpha [C_1^2 \cosh ^2(v) \sin 2 \phi -4 C_1 q \sinh (v) \sin ^2\phi \nonumber\\
&&-2 \phi ''(v)+\tanh ^2(v) \sin 2 \phi -\text{sech}^2(v) 
\sin 2 \phi ]\nonumber\\
&&+\alpha '(v) \sin 2 \alpha [C_1 q \cosh (v)+\phi '(v)] = 0.
\end{eqnarray}
and,
\begin{eqnarray}
&&\frac{1}{4} (\sin 2\alpha (2 {C_1}^2 \cosh ^2(v) \cos ^2 \phi 
-2 {C_1} q \sinh (v) \sin 2\phi \nonumber\\
&&-2 \text{sech}^2(v) \cos ^2\phi+\tanh ^2(v) \cos 2\phi+\tanh ^2(v))
\nonumber\\
&&-4 {C_1} q \cosh (v) \sin 2 \alpha  \phi '(v)-4 \alpha ''(v)-2 \sin 2 \alpha  \phi '(v)^2) = 0.\nonumber\\
\end{eqnarray}
We impose similar boundary conditions as before: 1) at the center of
the catenoid, at $v=0$, $\alpha (0)=\phi (0)=0$, consistent with 
$\phi,\alpha$ being odd fucntions, and 2) the upper 
and lower edges of the catenoid ($v=\pm v_f$) are torque free,
imposing $\frac{\partial \cal F}{\partial \phi'(r)}=0$ and 
$\frac{\partial \cal F}{\partial \alpha'(r)}=0$. That yields 
\begin{eqnarray}
\cos ^2\alpha [2C_1 q \cosh v_f
+(1-K_{24}) \tanh v_f \sin 2 \phi +2 \phi ']=0, \label{eq.cat.bc1}\\
\mbox{and},\;\;\;\;4\;\alpha ' [K_{24} (C_1 - \text{sech}\; {v_f}) \cos 2\alpha + C_1] + 
\tanh v_f \sin 2\alpha \nonumber\\
\times [2C_1 - K_{24} \text{sech}\; v_f + 2C_1 (K_{24}-1) \sin ^2 \phi  ]=0 \label{eq.cat.bc2}
\end{eqnarray}
With these boundary conditions the Euler equation fail to produce any 
$\alpha\neq 0$ solution, for catenoids with experimentally observed 
\cite{apsabstract} aspect ratios $v_f/C_1\sim 1.3$ (with $v_f\sim 2.8\mu m$ and $C_1\sim 2.1 \mu m$). In Fig\ref{fig:Cat1}a-c  we show the variation 
of $\phi,\alpha$ with $v$, at different aspect ratios $v_f/C_1$. 
There are two types of solutions, which exhibit small (up to $15^o$)
and large (up to $150^o$) values of $\phi$. Fig\ref{fig:Cat1}-d shows 
a phase diagram where large
$\phi$ solutions are confined in a narrow strip bordered by small $\phi$
solutions. In Fig\ref{fig:Cat2} we show the nematic arrangements corresponding to,  a) small $\phi$, and b) large $\phi$ solutions. In 
'b' the director is normal to the surface at $v=0$. As $|v|$ increases
it gradually turns parallel to the surface and later again turns away
from the surface towards surface normal. Due to fixed intrinsic chirality 
the director maintains an uniform direction of rotation (clockwise here)
towards increasing $v$.  

Analysis of the  boundary conditions 
Eqs.\ref{eq.cat.bc1},\ref{eq.cat.bc2} reveals that in order to realise 
$\alpha=0$ at the edges (as measured in experiments) $\alpha'$ 
also has to be zero there. Numerical solutions in Fig\ref{fig:Cat1}a-c (using Mathematica) does not admit such solutions. We therefore 
explore another boundary condition where $\alpha$ will be set to zero
at the edges, dropping  the torque free BC (Eq.\ref{eq.cat.bc2}) with 
respect to $\alpha$. The other torque free BC with respect to $\phi$ 
(Eq.\ref{eq.cat.bc1}) is retained. The only justification for this 
$\alpha(\pm v_f)=0$ BC is phenomenology, as we are not aware of any 
other physical reasoning. Similar BC has been applied before at the 
boundary of 3D structures \cite{softgij}.   

With this BC, solutions with nonzero $\alpha$ arise for catenoids 
with small aspect ratios $v_f/C_1 \leq 1$. Fig\ref{fig:Cat3}a-g show 
solutions for $\phi (v),\alpha (v)$ for different aspect ratios.
Three different type of solutions arise (described in the caption
of Fig.\ref{fig:Cat3}). Subplot (h) shows a phase diagram populated
by these solutions. Broadly,  $\alpha\neq 0$ solutions appear for
large $\lambda_p$ and  helicoids with small
aspect ratios. In these solutions $|\phi|$ 
keep increasing towards the edges upto $\sim 140^o$. This feature qualitatively matches  with experiments, however in experiments 
$\phi$ reaches upto $90^o$.  At small aspect ratios 
(Fig\ref{fig:Cat3}f,g), $\phi$ shows abrupt jumps along the 
non-azimuthal direction $v$, which is  reminiscent of domain 
boundaries. In Fig.\ref{fig:Cat4} two representative nematic 
arrangements corresponding to $\alpha\neq 0$ solutions are shown. 
 
For all the $\phi,\alpha$ plots (for disk, helicoid and catenoid) 
we used representaive values $q=0.5 \mu m^{-1}, \lambda_p=0.5 \mu 
m, K_{24}/K=1$. Changing these values merely shift the 
the phase boundaries but do not change the qualitative
nature of the solutions. Further, since we used nondimensional
variable $q\lambda_p$, change of $\lambda_p$ at fixed $q$ will
have the same effect as changing $q$ keeping $\lambda_p$ fixed.

In summary, we examined how rod like molecules preferentially arrange 
on a curved collodal membranes having the shape of minimal surfaces. 
In particuler, we focused on possible departure of the rods (by angle 
$\alpha$) from the plane formed by the surface normal and the azimuthal tangent vector. This was motivated by recent measurements on catenoid 
shaped colloidal membranes \cite{apsabstract}.
We also showed how changing boundary conditions change the nature
of the solutions and allow comparison of theory to experimental
measurements. It was interesting that for disks and helicoids,
torque free BC with respect to both $\phi$ and $\alpha$ reproduce
experimental results, however for catenoids a more phenomenological
BC $\alpha(\pm v_f)=0$ had to be adopted to produce nonzero $\alpha$. 
Our phase diagrams summarise how the nature of the solutions 
depend on material property ($\lambda_p$) and the geometry of the
surface (via aspect ratio). 
However, we note that a complete solution to this problem will require simultaneous minimization of the membrane geometry and the director configuration, which requires optimization in a bigger parameter space.  
Although colloidal membranes with minimal surfaces have been realised 
in the laboratory, measuring precise director orientations remain challenging. Our theoretical predictions will motivate this effort.

\section{Acknowledgements}
AA and AS acknowledge Science and Engineering Research Board (SERB), India
(Project No. CRG/2019/005944), and IRCC-IIT Bombay, India for financial support.
AB thanks UGC-India for financial support. PS acknowledges SERB, India (Project No. CRG/2019/000855) for funding.

\noindent $^*$ asain@phy.iitb.ac.in

\bibliography{Ref_papers}

\begin{thebibliography}{17}
\expandafter\ifx\csname natexlab\endcsname\relax\def\natexlab#1{#1}\fi
\expandafter\ifx\csname bibnamefont\endcsname\relax
  \def\bibnamefont#1{#1}\fi
\expandafter\ifx\csname bibfnamefont\endcsname\relax
  \def\bibfnamefont#1{#1}\fi
\expandafter\ifx\csname citenamefont\endcsname\relax
  \def\citenamefont#1{#1}\fi
\expandafter\ifx\csname url\endcsname\relax
  \def\url#1{\texttt{#1}}\fi
\expandafter\ifx\csname urlprefix\endcsname\relax\def\urlprefix{URL }\fi
\providecommand{\bibinfo}[2]{#2}
\providecommand{\eprint}[2][]{\url{#2}}

\bibitem[{\citenamefont{Aggeli et~al.}(2001)\citenamefont{Aggeli, Nyrkova,
  Bell, Harding, Carrick, McLeish, Semenov, and Boden}}]{chiralstrucPNAS2001}
\bibinfo{author}{\bibfnamefont{A.}~\bibnamefont{Aggeli}},
  \bibinfo{author}{\bibfnamefont{I.~A.} \bibnamefont{Nyrkova}},
  \bibinfo{author}{\bibfnamefont{M.}~\bibnamefont{Bell}},
  \bibinfo{author}{\bibfnamefont{R.}~\bibnamefont{Harding}},
  \bibinfo{author}{\bibfnamefont{L.}~\bibnamefont{Carrick}},
  \bibinfo{author}{\bibfnamefont{T.~C.} \bibnamefont{McLeish}},
  \bibinfo{author}{\bibfnamefont{A.~N.} \bibnamefont{Semenov}},
  \bibnamefont{and} \bibinfo{author}{\bibfnamefont{N.}~\bibnamefont{Boden}},
  \bibinfo{journal}{Proceedings of the National Academy of Sciences}
  \textbf{\bibinfo{volume}{98}}, \bibinfo{pages}{11857} (\bibinfo{year}{2001}).

\bibitem[{\citenamefont{Barry et~al.}(2009{\natexlab{a}})\citenamefont{Barry,
  Beller, and Dogic}}]{memdogic2009}
\bibinfo{author}{\bibfnamefont{E.}~\bibnamefont{Barry}},
  \bibinfo{author}{\bibfnamefont{D.}~\bibnamefont{Beller}}, \bibnamefont{and}
  \bibinfo{author}{\bibfnamefont{Z.}~\bibnamefont{Dogic}},
  \bibinfo{journal}{Soft Matter} \textbf{\bibinfo{volume}{5}},
  \bibinfo{pages}{2563} (\bibinfo{year}{2009}{\natexlab{a}}).

\bibitem[{\citenamefont{Barry and Dogic}(2010)}]{memdogic2010}
\bibinfo{author}{\bibfnamefont{E.}~\bibnamefont{Barry}} \bibnamefont{and}
  \bibinfo{author}{\bibfnamefont{Z.}~\bibnamefont{Dogic}},
  \bibinfo{journal}{Proceedings of the National Academy of Sciences}
  \textbf{\bibinfo{volume}{107}}, \bibinfo{pages}{10348}
  (\bibinfo{year}{2010}).

\bibitem[{\citenamefont{Saikia et~al.}(2017)\citenamefont{Saikia, Sarkar,
  Thomas, Raghunathan, Sain, and Sharma}}]{prerna}
\bibinfo{author}{\bibfnamefont{L.}~\bibnamefont{Saikia}},
  \bibinfo{author}{\bibfnamefont{T.}~\bibnamefont{Sarkar}},
  \bibinfo{author}{\bibfnamefont{M.}~\bibnamefont{Thomas}},
  \bibinfo{author}{\bibfnamefont{V.}~\bibnamefont{Raghunathan}},
  \bibinfo{author}{\bibfnamefont{A.}~\bibnamefont{Sain}}, \bibnamefont{and}
  \bibinfo{author}{\bibfnamefont{P.}~\bibnamefont{Sharma}},
  \bibinfo{journal}{Nature communications} \textbf{\bibinfo{volume}{8}},
  \bibinfo{pages}{1160} (\bibinfo{year}{2017}).

\bibitem[{\citenamefont{Gibaud et~al.}(2012)\citenamefont{Gibaud, Barry,
  Zakhary, Henglin, Ward, Yang, Berciu, Oldenbourg, Hagan, Nicastro
  et~al.}}]{Gibaud}
\bibinfo{author}{\bibfnamefont{T.}~\bibnamefont{Gibaud}},
  \bibinfo{author}{\bibfnamefont{E.}~\bibnamefont{Barry}},
  \bibinfo{author}{\bibfnamefont{M.~J.} \bibnamefont{Zakhary}},
  \bibinfo{author}{\bibfnamefont{M.}~\bibnamefont{Henglin}},
  \bibinfo{author}{\bibfnamefont{A.}~\bibnamefont{Ward}},
  \bibinfo{author}{\bibfnamefont{Y.}~\bibnamefont{Yang}},
  \bibinfo{author}{\bibfnamefont{C.}~\bibnamefont{Berciu}},
  \bibinfo{author}{\bibfnamefont{R.}~\bibnamefont{Oldenbourg}},
  \bibinfo{author}{\bibfnamefont{M.~F.} \bibnamefont{Hagan}},
  \bibinfo{author}{\bibfnamefont{D.}~\bibnamefont{Nicastro}},
  \bibnamefont{et~al.}, \bibinfo{journal}{Nature}
  \textbf{\bibinfo{volume}{481}}, \bibinfo{pages}{348} (\bibinfo{year}{2012}).

\bibitem[{\citenamefont{{Balchunas} et~al.}(2017)\citenamefont{{Balchunas},
  {Sharma}, and {Dogic}}}]{apsabstract}
\bibinfo{author}{\bibfnamefont{A.}~\bibnamefont{{Balchunas}}},
  \bibinfo{author}{\bibfnamefont{P.}~\bibnamefont{{Sharma}}}, \bibnamefont{and}
  \bibinfo{author}{\bibfnamefont{Z.}~\bibnamefont{{Dogic}}}, in
  \emph{\bibinfo{booktitle}{APS March Meeting Abstracts}}
  (\bibinfo{year}{2017}), vol. \bibinfo{volume}{2017} of
  \emph{\bibinfo{series}{APS Meeting Abstracts}}, p. \bibinfo{pages}{C17.011}.

\bibitem[{\citenamefont{Balchunas et~al.}(2020)\citenamefont{Balchunas, Jia,
  Zakhary, Robaszewski, Gibaud, Dogic, Pelcovits, and Powers}}]{powers1}
\bibinfo{author}{\bibfnamefont{A.}~\bibnamefont{Balchunas}},
  \bibinfo{author}{\bibfnamefont{L.~L.} \bibnamefont{Jia}},
  \bibinfo{author}{\bibfnamefont{M.~J.} \bibnamefont{Zakhary}},
  \bibinfo{author}{\bibfnamefont{J.}~\bibnamefont{Robaszewski}},
  \bibinfo{author}{\bibfnamefont{T.}~\bibnamefont{Gibaud}},
  \bibinfo{author}{\bibfnamefont{Z.}~\bibnamefont{Dogic}},
  \bibinfo{author}{\bibfnamefont{R.~A.} \bibnamefont{Pelcovits}},
  \bibnamefont{and} \bibinfo{author}{\bibfnamefont{T.~R.}
  \bibnamefont{Powers}}, \bibinfo{journal}{Physical Review Letters}
  \textbf{\bibinfo{volume}{125}}, \bibinfo{pages}{018002}
  (\bibinfo{year}{2020}).

\bibitem[{\citenamefont{Gibaud et~al.}(2017)\citenamefont{Gibaud, Kaplan,
  Sharma, Zakhary, Ward, Oldenbourg, Meyer, Kamien, Powers, and
  Dogic}}]{gibaud2017}
\bibinfo{author}{\bibfnamefont{T.}~\bibnamefont{Gibaud}},
  \bibinfo{author}{\bibfnamefont{C.~N.} \bibnamefont{Kaplan}},
  \bibinfo{author}{\bibfnamefont{P.}~\bibnamefont{Sharma}},
  \bibinfo{author}{\bibfnamefont{M.~J.} \bibnamefont{Zakhary}},
  \bibinfo{author}{\bibfnamefont{A.}~\bibnamefont{Ward}},
  \bibinfo{author}{\bibfnamefont{R.}~\bibnamefont{Oldenbourg}},
  \bibinfo{author}{\bibfnamefont{R.~B.} \bibnamefont{Meyer}},
  \bibinfo{author}{\bibfnamefont{R.~D.} \bibnamefont{Kamien}},
  \bibinfo{author}{\bibfnamefont{T.~R.} \bibnamefont{Powers}},
  \bibnamefont{and} \bibinfo{author}{\bibfnamefont{Z.}~\bibnamefont{Dogic}},
  \bibinfo{journal}{Proceedings of the National Academy of Sciences}
  \textbf{\bibinfo{volume}{114}}, \bibinfo{pages}{E3376}
  (\bibinfo{year}{2017}).

\bibitem[{\citenamefont{Barry et~al.}(2009{\natexlab{b}})\citenamefont{Barry,
  Dogic, Meyer, Pelcovits, and Oldenbourg}}]{measured09}
\bibinfo{author}{\bibfnamefont{E.}~\bibnamefont{Barry}},
  \bibinfo{author}{\bibfnamefont{Z.}~\bibnamefont{Dogic}},
  \bibinfo{author}{\bibfnamefont{R.~B.} \bibnamefont{Meyer}},
  \bibinfo{author}{\bibfnamefont{R.~A.} \bibnamefont{Pelcovits}},
  \bibnamefont{and}
  \bibinfo{author}{\bibfnamefont{R.}~\bibnamefont{Oldenbourg}},
  \bibinfo{journal}{The Journal of Physical Chemistry B}
  \textbf{\bibinfo{volume}{113}}, \bibinfo{pages}{3910}
  (\bibinfo{year}{2009}{\natexlab{b}}).

\bibitem[{\citenamefont{Pelcovits and Meyer}(2009)}]{meyer}
\bibinfo{author}{\bibfnamefont{R.~A.} \bibnamefont{Pelcovits}}
  \bibnamefont{and} \bibinfo{author}{\bibfnamefont{R.~B.} \bibnamefont{Meyer}},
  \bibinfo{journal}{Liquid Crystals} \textbf{\bibinfo{volume}{36}},
  \bibinfo{pages}{1157} (\bibinfo{year}{2009}).

\bibitem[{\citenamefont{Selinger and Schnur}(1993)}]{selingerPRL}
\bibinfo{author}{\bibfnamefont{J.~V.} \bibnamefont{Selinger}} \bibnamefont{and}
  \bibinfo{author}{\bibfnamefont{J.~M.} \bibnamefont{Schnur}},
  \bibinfo{journal}{Physical review letters} \textbf{\bibinfo{volume}{71}},
  \bibinfo{pages}{4091} (\bibinfo{year}{1993}).

\bibitem[{\citenamefont{Helfrich and Prost}(1988)}]{HFterm}
\bibinfo{author}{\bibfnamefont{W.}~\bibnamefont{Helfrich}} \bibnamefont{and}
  \bibinfo{author}{\bibfnamefont{J.}~\bibnamefont{Prost}},
  \bibinfo{journal}{Physical Review A} \textbf{\bibinfo{volume}{38}},
  \bibinfo{pages}{3065} (\bibinfo{year}{1988}).

\bibitem[{\citenamefont{Schmidt}(1990)}]{k24-90}
\bibinfo{author}{\bibfnamefont{V.~H.} \bibnamefont{Schmidt}},
  \bibinfo{journal}{Physical review letters} \textbf{\bibinfo{volume}{64}},
  \bibinfo{pages}{535} (\bibinfo{year}{1990}).

\bibitem[{\citenamefont{Allender et~al.}(1991)\citenamefont{Allender, Crawford,
  and Doane}}]{k24-91}
\bibinfo{author}{\bibfnamefont{D.~W.} \bibnamefont{Allender}},
  \bibinfo{author}{\bibfnamefont{G.}~\bibnamefont{Crawford}}, \bibnamefont{and}
  \bibinfo{author}{\bibfnamefont{J.}~\bibnamefont{Doane}},
  \bibinfo{journal}{Physical review letters} \textbf{\bibinfo{volume}{67}},
  \bibinfo{pages}{1442} (\bibinfo{year}{1991}).

\bibitem[{\citenamefont{McInerney et~al.}(2019)\citenamefont{McInerney, Ellis,
  Rocklin, Fernandez-Nieves, and Matsumoto}}]{softgij}
\bibinfo{author}{\bibfnamefont{J.~P.} \bibnamefont{McInerney}},
  \bibinfo{author}{\bibfnamefont{P.~W.} \bibnamefont{Ellis}},
  \bibinfo{author}{\bibfnamefont{D.~Z.} \bibnamefont{Rocklin}},
  \bibinfo{author}{\bibfnamefont{A.}~\bibnamefont{Fernandez-Nieves}},
  \bibnamefont{and} \bibinfo{author}{\bibfnamefont{E.~A.}
  \bibnamefont{Matsumoto}}, \bibinfo{journal}{Soft matter}
  \textbf{\bibinfo{volume}{15}}, \bibinfo{pages}{1210} (\bibinfo{year}{2019}).

\bibitem[{\citenamefont{Kang and Lubensky}(2017)}]{lubenskyRaft}
\bibinfo{author}{\bibfnamefont{L.}~\bibnamefont{Kang}} \bibnamefont{and}
  \bibinfo{author}{\bibfnamefont{T.~C.} \bibnamefont{Lubensky}},
  \bibinfo{journal}{Proceedings of the National Academy of Sciences}
  \textbf{\bibinfo{volume}{114}}, \bibinfo{pages}{E19} (\bibinfo{year}{2017}).

\bibitem[{\citenamefont{Kaplan et~al.}(2010)\citenamefont{Kaplan, Tu,
  Pelcovits, and Meyer}}]{kaplan2010theory}
\bibinfo{author}{\bibfnamefont{C.~N.} \bibnamefont{Kaplan}},
  \bibinfo{author}{\bibfnamefont{H.}~\bibnamefont{Tu}},
  \bibinfo{author}{\bibfnamefont{R.~A.} \bibnamefont{Pelcovits}},
  \bibnamefont{and} \bibinfo{author}{\bibfnamefont{R.~B.} \bibnamefont{Meyer}},
  \bibinfo{journal}{Physical Review E} \textbf{\bibinfo{volume}{82}},
  \bibinfo{pages}{021701} (\bibinfo{year}{2010}).

\end{thebibliography}
\end{document}